\documentclass[12pt]{amsart}

\usepackage{graphicx}
\newcommand{\R}{\mathbb R}

\newcommand{\x}{\mathbf x}
\newcommand{\y}{\mathbf y}
\newcommand{\p}{\mathbf p}
\newcommand{\jj}{\mathbf j}

\newcommand{\grad}{\mathop{\rm grad}\nolimits}

\newcommand{\CC}{\mathbb C}
\newcommand{\PV}{\mathcal{PV}}
\newcommand{\Imm}{\mathop{\rm Im}\nolimits}
\newcommand{\Sp}{\mathop{\rm Sp}\nolimits}
\newcommand{\GL}{\mathop{\rm GL}\nolimits}
\newcommand{\Mp}{\mathop{\rm Mp}\nolimits}
\newcommand{\SG}{\mathcal{SG}}
\newcommand{\calC}{\mathcal C}

\newcommand{\const}{\mathop{\rm const}\nolimits}
\newcommand{\Heis}{\mathcal{H}}
\begin{document}

\title[Maslov's complex germ and the Weyl--Moyal algebra]
{Maslov's complex germ and the Weyl--Moyal algebra
in quantum mechanics and in quantum field theory}
\author{A. V. Stoyanovsky}
\email{stoyan@mccme.ru}

\begin{abstract}
The paper is a survey of some author's results related with
the Maslov--Shvedov method of complex germ and with quantum
field theory.
The main idea is that many results of the method of complex germ
and of perturbative quantum field theory
can be made more simple and natural if instead of the algebra of
(pseudo)differential operators one uses the Weyl algebra
(operators with Weyl symbols) with the Moyal $*$-product.
\end{abstract}

\maketitle

\tableofcontents

\section*{Introduction}

This paper is a survey of some author's results related
with the Maslov--Shvedov method of complex germ [2]
and with perturbative quantum field theory.
These results are, shortly, the following.

Firstly, the results from the theory of quantum mechanical Schrod\-ing\-er equation.
The Maslov--Shvedov method makes it possible to give a simple exposition
of the method of canonical operator on a Lagrangian manifold with complex germ [2--5]
of asymptotic solution of the Cauchy problem for the Schrodinger equation.
It turns out that many Maslov--Shvedov's formulas become more simple and get a
natural mathematical interpretation if in the Schrodinger equation one chooses
the Weyl (symmetric) ordering of multiplication and differentiation operators.
In particular, the transport equation describes the transport of half-forms
along the classical trajectory, and the transport of Maslov--Shvedov wave packets
is given by the action of the operator of the Weil representation of the metaplectic group
corresponding to the tangent symplectic transformation to the Hamiltonian flow.
Besides that, one obtains a simple and natural definition of the Maslov index modulo 4,
related with the complex germ method at a point, which seemingly
did not appear in the literature. A closed exposition of these results
is given in \S1.

Secondly, the results from perturbative quantum field theory. Here
introducing the infinite dimensional analog of the Weyl algebra
allows one not only to interpret the quantum field theory Maslov--Shvedov
complex germ obtained in [2] by complicated computations
(or, more precisely, to obtain a result close to that result of [2]),
but, generally, it allows one to give an exposition of the main results of
perturbative quantum field theory not using subtraction of infinities from
the Hamiltonian of a free field and normal ordering of operators.
The Weyl algebra plays the role of an extended algebra of operators
in the Fock space, where the $*$-product corresponds to
the composition of operators, while the usual commutative product of functions
corresponds to normally ordered product of operators. This result seems
very important for understanding free and perturbative quantum field theory.
A closed exposition of these results on the simplest example of the
$\varphi^4$ model of quantum field theory in four-dimensional space-time
is contained in \S2.

The author is grateful to V.~V.~Dolotin, Yu.~A.~Neretin, and I.~V.~Tyu\-tin
for helpful discussions.

\section{Complex germ and the Weyl algebra in quantum mechanics}

\subsection{The Schrodinger equation}

The Schrodinger equation reads
\begin{equation}
-ih\frac{\partial\psi}{\partial t}+
\hat H(t,-ih\frac{\partial}{\partial q_1},\ldots,
-ih\frac{\partial}{\partial q_n},q_1,\ldots,q_n)\psi=0.
\end{equation}
Here $\psi(t,q_1,\ldots,q_n)$ is an unknown complex valued function (the wave function
of a quantum mechanical system), $H(t,p,q)$ is the Hamiltonian of the corresponding
system of classical mechanics, $q=(q_1$, $\ldots$, $q_n)$,
$p=(p_1$, $\ldots$, $p_n)$. This equation is written in such a way that after substitution,
instead of $\psi$, of the {\it quasiclassical asymptotics}
\begin{equation}
\psi=a(t,q_1,\ldots,q_n)e^{i\frac{S(t,q_1,\ldots,q_n)}{h}}
\end{equation}
($a$ and $S$ are real functions varying very slowly when compared with the number $h$),
in the principal approximation as $h\to 0$
({\it the quasiclassical limit}) one would obtain the {\it Hamilton--Jacobi equation}
\begin{equation}
\frac{\partial S}{\partial t}+
H(t,\frac{\partial S}{\partial q_1},\ldots,
\frac{\partial S}{\partial q_n},q_1,\ldots,q_n)=0
\end{equation}
for the function $S$.

The form of the Schrodinger equation (1) has some ambiguity in the general case. It is
related with the fact that the operators of multiplication by a function of $q_i$
and differentiation with respect to $q_i$, in general, do not commute.
Hence one should, in general, choose an ordering of these operators in the quantum
Hamiltonian $\hat H$. In the case of a standard mechanical system without
constraints, in which
\begin{equation}
H(t,p,q)=\sum\frac{p_i^2}{2m_i}+U(t,q),
\end{equation}
there is no such ambiguity. However, more general and deep considerations require
to overcome this ambiguity.  The usual way is to put the operators of differentiation
with respect to $q_i$ to the right of operators of multiplication by a function.

\subsection{Asymptotic Cauchy problem}

Let us pose the asymptotic Cauchy problem for the Schrodinger equation (1):
take oscillating initial data
\begin{equation}
\psi_0(q_1,\ldots,q_n)=a_0(q)e^{iS_0(q)/h},
\end{equation}
and let us look for an oscillating function $\psi(t,q)$ of the form (2),
which turns into $\psi_0$ for
$t=0$ and satisfies equation (1) up to $o(h)$. To that end, two equations should hold:
the Hamilton--Jacobi equation (3) and the {\it transport equation}, obtained
by equating coefficients before $h$ in the Schrodinger equation, into which the quasiclassical
solution (2) is substituted.
It is not difficult to see that, in the case when all the operators
$\frac{\partial}{\partial q_i}$ are put to the right
of the operators $q_i$, the transport equation reads
\begin{equation}
\begin{aligned}{}
\frac{\partial a}{\partial t}&+\sum_i
\frac{\partial a}{\partial q_i}H_{p_i}(t,q_1,\ldots,q_n,
\frac{\partial S}{\partial q_1},\ldots,\frac{\partial S}{\partial q_n})\\
&+\frac a2\sum_{i,j}H_{p_ip_j}(t,q_1,\ldots,q_n,
\frac{\partial S}{\partial q_1},\ldots,\frac{\partial S}{\partial q_n})
\frac{\partial^2S}{\partial q_i\partial q_j}=0.
\end{aligned}
\end{equation}
Thus, our asymptotic Cauchy problem has been reduced to the Cauchy problem for the system
of equations (3,6). It is well known that the Cauchy problem for the Hamilton--Jacobi equation
amounts to integration of the system of ordinary {\it characteristic}
Hamilton differential equations
\begin{equation}
\frac{dq_i}{dt}=\frac{\partial H}{\partial p_i},\ \
\frac{dp_i}{dt}=-\frac{\partial H}{\partial q_i}.
\end{equation}
(See, for example, [6], Ch.~4, or [1], Ch.~2.) Assume that this problem is already solved.
How can one find the function $a(t,q)$? The transport equation is an ordinary differential
equation for the function $a$, giving its behavior along the trajectories of
the system of ordinary differential equations (7), where $p_i=\frac{\partial S}{\partial q_i}$.
Considering particular cases (4) from standard quantum mechanics gives a solution of the
form
\begin{equation}
a(t,q(t))=a_0(q(0))\frac1{\sqrt{\det\left(\frac{\partial q_i(t)}{\partial q_j(0)}\right)}},
\end{equation}
where $(p_i(t),q_i(t))$ is a characteristic, i.~e., a solution of the Hamilton equations.
Indeed, let us differentiate $\frac{\partial q_i(t)}{\partial q_j(0)}$ with respect to time:
\begin{equation}
\begin{aligned}{}
&\frac d{dt}\frac{\partial q_i(t,q(0))}{\partial q_j(0)}
=\frac{\partial H_{p_i}}{\partial q_j(0)}=\sum H_{p_ip_k}\frac{\partial p_k}{\partial q_j(0)}
+\sum H_{p_iq_l}\frac{\partial q_l}{\partial q_j(0)}\\
&=\sum H_{p_ip_k}\frac{\partial^2S}{\partial q_k\partial q_l}\frac{\partial q_l}{\partial q_j(0)}
+\sum H_{p_iq_l}\frac{\partial q_l}{\partial q_j(0)},
\end{aligned}
\end{equation}
whence
\begin{equation}
\begin{aligned}{}
\frac{da(t,q(t))}{dt}&=\frac{\partial a}{\partial t}
+\sum H_{p_i}\frac{\partial a}{\partial q_i}\\
&=-\frac12a\left(\sum H_{p_ip_k}\frac{\partial^2S}{\partial q_i\partial q_k}
+\sum H_{p_iq_i}\right).
\end{aligned}
\end{equation}
This equation differs from the transport equation (6) by the term
$$
\frac12a\sum H_{p_iq_i},
$$
which is zero in the standard case (4). It turns out that here the point is the
ordering of non-commuting operators $-ih\frac{\partial}{\partial q_i}$ and $q_i$
in the Hamiltonian. If we choose a ``right'' ordering (recall that above we have
arbitrarily put $-ih\frac{\partial}{\partial q_i}$ to the right of $q_j$), then
the difference between equations (10)
and (6) will disappear. The choice of a ``right'' ordering is the subject
of the following two subsections.

\subsection{The Weil representation}

As a simplest example consider the Hamiltonian $H=pq$ with $n=1$. For it, equation
(10) requires the choice
\begin{equation}
\hat H=-ihq\frac\partial{\partial q}-\frac{ih}2=-\frac{ih}2\left(q\cdot\frac\partial{\partial q}
+\frac\partial{\partial q}\cdot q\right)
\end{equation}
instead of $-ihq\frac\partial{\partial q}$ chosen above. That is, operators
$-ih\frac\partial{\partial q_i}$ and $q_i$ should belong to the Hamiltonian
symmetrically, without a prescription what stands to the right and what stands to the left.
For formalization of these requirements, we need some information on the symplectic group.

Consider the general quantum quadratic Hamiltonians
\begin{equation}
\hat H=\sum\frac12 a_{jk}q_jq_k-\frac{ih}2 b_{jk}
\left(q_j\frac\partial{\partial q_k}+\frac\partial{\partial q_k}q_j\right)
-\frac{h^2}2c_{jk}\frac\partial{\partial q_j}\frac\partial{\partial q_k}.
\end{equation}
Here $a_{jk}$ and $c_{jk}$ are symmetric real matrices, $b_{jk}$ is an arbitrary real
matrix. In the Cauchy problem let us put the initial condition
\begin{equation}
\psi_0=\psi_{Z,p_0}=\exp\frac ih\left(\frac12\sum Z_{ij}q_iq_j+\sum p_0^jq_j\right).
\end{equation}
Here $Z=(Z_{ij})$ is a symmetric complex matrix, $p_0$ is a real vector.
Assume that the matrix $Z$ has positive definite imaginary part, then function
(13) rapidly decreases at infinity ({\it Gaussian wave packet}). It turns out that,
as it is not difficult to check by a direct computation, formula (8)
gives in this case not only asymptotic but exact solution of the Cauchy problem:
\begin{equation}
\psi(t,q)=\frac{\psi_{(AZ+B)(CZ+D)^{-1},((CZ+D)^T)^{-1}p_0}}{\sqrt{\det(CZ+D)}}
e^{-\frac i{2h}p_0^T(CZ+D)^{-1}Cp_0}.
\end{equation}
Here the sign $T$ denotes transposing;
$A,B,C,D$ are matrices which can be found in the following way. The characteristics
equations read
\begin{equation}
\begin{aligned}{}
dq_i/dt&=H_{p_i}=\sum_j b_{ji}q_j+\sum_j c_{ij}p_j,\\
dp_i/dt&=-H_{q_i}=-\sum_j a_{ij}q_j-\sum_j b_{ij}p_j
\end{aligned}
\end{equation}
with the initial conditions
\begin{equation}
p_i(0)=\sum Z_{ij}q_j(0)+p_0^i.
\end{equation}
These are linear equations, hence, the evolution operator at the time $t$ is a
linear operator
\begin{equation}
\exp t\left(\begin{array}{cc}
-b&-a\\
c&b^T
\end{array}
\right)=\left(\begin{array}{cc}
A&B\\
C&D
\end{array}
\right).
\end{equation}
In addition, this operator preserves the Poisson bracket of any two functions, as any
evolution operator of the canonical Hamilton equations, i.~e., it preserves the
bivector field
\begin{equation}
\eta=\sum\frac\partial{\partial p_i}\wedge\frac\partial{\partial q_i}.
\end{equation}
In the language of matrices this condition means that
\begin{equation}
\left(\begin{array}{cc}
A&B\\
C&D
\end{array}
\right)\left(\begin{array}{cc}
0&E\\
-E&0
\end{array}
\right)\left(\begin{array}{cc}
A^T&C^T\\
B^T&D^T
\end{array}
\right)=\left(\begin{array}{cc}
0&E\\
-E&0
\end{array}
\right),
\end{equation}
where $E$ is the unit matrix. Such linear operators are called symplectic,
they form the symplectic group $\Sp(2n,\R)$.
The set of symmetric complex matrices $Z$ with positive definite imaginary part
is called the {\it Siegel upper half-plane} (cf. [7]);
denote it by $\SG$. The group $\Sp(2n,\R)$
acts on the half-plane $\SG$ by the formula
\begin{equation}
Z\mapsto(AZ+B)(CZ+D)^{-1}.
\end{equation}
Equivalently, this action can be defined by means of evolution at the time $t$
of the matrix Riccati equation
\begin{equation}
\dot Z+ZcZ+bZ+Zb^T+a=0,
\end{equation}
which is obtained by differentiating the action (20) with respect to $t$.

Formulas (13), (14) define the action of the group $\Sp(2n,\R)$ on the set of
Gaussian wave packets. But this action is two-valued: to make it single-valued,
one must choose one of the two continuous branches of the square root from $\det(CZ+D)\ne 0$,
$Z\in\SG$.
Hence a two-fold covering of the group $\Sp(2n,\R)$ has a single-valued action on the
set of Gaussian wave packets. This covering is called the {\it metaplectic group};
denote it by $\Mp(2n,\R)$.

It turns out that the action of the group $\Mp(2n,\R)$ on the set of Gaussian
wave packets is uniquely extended by continuousness to the action on the Schwartz space
$S=S(\R^n)$
of complex valued smooth functions $\psi(q_1,\ldots,q_n)$ rapidly decreasing at infinity,
and also to unitary action on the space $L_2(\R^n)$ of square integrable functions
and to the action on the dual to $S$ space $S'=S'(\R^n)$ of tempered distributions.
Also the product of matrices yields the composition of operators. This representation
of the group $\Mp(2n,\R)$ is called the {\it Weil representation}, cf. [8,9].

The Weil representation is uniquely, up to a constant factor, characterized by the
following property. Conjugation by an operator $U$ corresponding to the matrix
$\left(\begin{array}{cc}
A&B\\
C&D
\end{array}
\right)$, preserves the $2n$-dimensional vector space
of operators with the basis
\begin{equation}
(q_1,\ldots,q_n,ih\frac{\partial}{\partial q_1},\ldots,
ih\frac{\partial}{\partial q_n})
\end{equation}
and acts on this space by the matrix
$\left(\begin{array}{cc}
A&B\\
C&D
\end{array}
\right)$. This is obtained by exponentiating from the fact that the
commutator with the Hamiltonian (12) also preserves this space, and acts on it by the
matrix $\left(\begin{array}{cc}
-b&-a\\
c&b^T
\end{array}
\right)$, up to the factor $ih$. This property implies the uniqueness of the
Weil representation as follows.
If $U'$ is another operator with the same property, then the operator $U'U^{-1}$
commutes with the operators $q_j$ and $ih\frac{\partial}{\partial q_j}$, and hence it is
multiplication by a constant, as it is not difficult to show.

In particular, the matrix
$\left(\begin{array}{cc}E&B\\0&E\end{array}\right)$ acts by multiplication by the function
$\exp\left(\frac i{2h}\sum_{j,k}B_{jk}x_jx_k\right)$;
the matrix $\left(\begin{array}{cc}A&0\\0&(A^T)^{-1}\end{array}\right)$
acts by composition of a linear change of coordinates given by the matrix $A$,
and multiplication by $\sqrt{\det A}$. Finally,
the matrix $\left(\begin{array}{cc}0&-E\\E&0\end{array}\right)$
acts (up to a constant factor)
by the Fourier transform:
\begin{equation}
(F_h\psi)(q)=\frac1{\sqrt{(2\pi h)^n}}
\int e^{-i\sum q_jy_j/h}\psi(y)dy.
\end{equation}
This transform exchanges the operators $q_j$ and $ih\frac\partial{\partial q_j}$
(up to sign); the square of this transform is the change of variables $q\to-q$.
In quantum mechanics the Fourier transform of the wave function is called its
momentum representation.

The above matrices generate the group $\Sp(2n,\R)$,
which gives a proof of existence of an action of the group $\Sp(2n,\R)$,
defined up to a factor, with the above described commutation relations
with the operators (22).

Note also that exponentiating of operators (22) yields an action of the so called
{\it Heisenberg group} $\Heis_n$ on the space of functions. Namely, the operator
$a_1q_1+\ldots+a_nq_n$ acts by multiplication by the function $e^{i\sum a_jq_j/h}$,
and the operator $ih\sum b_i\frac{\partial}{\partial q_i}$ acts by the change of
variables $q_i\to q_i+b_i$. The multiplication in the group $\Heis_n$ is defined with the
help of the formula
\begin{equation}
e^{\hat a}e^{\hat b}=e^{\hat a+\hat b}\cdot e^{c/2},\ \ \hat a\hat b-\hat b\hat a=c
\end{equation}
(here $c$ is a number).
This action is compatible with the action of the group $\Mp(2n,\R)$ in an obvious sense,
so that the space of functions has an action of the {\it semidirect product}
of the groups $\Mp(2n,\R)$ and $\Heis_n$. This group is a central extension of the
affine symplectic group (i.~e., the semidirect product of the group $\Sp(2n,\R)$
and the group of parallel translations in the space $\R^{2n}$) with the help of the circle.
Let us denote the action of an element $g$ of any of these groups (possibly
defined up to a factor) on the space of functions by the symbol
$$
U=\rho(g).
$$

\subsection{The Weyl calculus}

This is a way to assign to a function
$\varphi(p_1$, $\ldots$, $p_n$, $q_1$, $\ldots$, $q_n)$ an operator
$\hat\varphi=\hat\varphi(p,q)$
on the space of functions
$\psi(q_1,\ldots,q_n)$, this correspondence being in accordance with the action of
the affine symplectic group, i.~e., for any element $g$ of this group we have
\begin{equation}
\rho(g)\hat\varphi(p,q)\rho(g)^{-1}=\widehat{g\varphi}(p,q),
\end{equation}
where $(g\varphi)(p,q)=\varphi(g^{-1}(p,q))$. The correspondence $\varphi\to\hat\varphi$
possesses also the following properties:

a) $((\sum a_iq_i+b_ip_i)^k)\widehat{\ }=(\sum a_iq_i-ihb_i\frac\partial{\partial q_i})^k$;

b) a real function $\varphi$ corresponds to a symmetric operator $\hat\varphi$,
i.~e. such that
$$
\int\overline\psi_1\cdot\hat\varphi\psi_2\,dq=\int\overline{\hat\varphi\psi_1}\cdot\psi_2\,dq
$$
for any rapidly decreasing functions $\psi_1$, $\psi_2$;

c) appropriate continuousness properties, into which we shall not go, see, for example,
Hormander's book [10].

Using these properties one can define the operator $\hat\varphi$ for a rather wide
class of functions $\varphi$. First of all, for polynomials $\varphi(p,q)$
the operator $\hat\varphi$
is defined uniquely from property~(a). For example,
\begin{equation}
(pq)\widehat{\ }=\frac12((p+q)^2-p^2-q^2)\widehat{\ }
=-ih\left(q\frac\partial{\partial q}+\frac12\right).
\end{equation}
Similarly, in algebra a way is known to express each polynomial
of $2n$ variables through powers of linear forms. For any homogeneous
polynomial of degree $k$
of $2n$ variables there exists a unique symmetric $k$-linear form of $2n$
variables (the {\it polarization} of the polynomial), which
gives this polynomial for coinciding arguments. A polylinear form is a tensor of rank $k$, i.~e.
an element of non-commutative algebra of $2n$ generators. Substituting instead of these
generators the operators $q_i$ and $-ih\frac\partial{\partial q_i}$, we obtain the
required operator. This operation is $\GL(2n,\R)$-invariant.

Further, we have
\begin{equation}
\left(\exp\frac ih(\sum a_ip_i+b_iq_i)\right)\widehat{\ }=
\exp\left(\sum a_i\frac\partial{\partial q_i}+\frac ihb_iq_i\right).
\end{equation}
Since many functions $\varphi$ can be expressed as superposition of exponents of linear forms
using the inverse Fourier transform:
\begin{equation}
\varphi(p,q)=\frac1{(2\pi h)^n}\int(F_h\varphi)(a,b)e^{\frac ih(\sum a_ip_i+b_iq_i)}\,dadb
\end{equation}
(see (23)), we obtain a way to find the required operator for a large class of
functions.
Let us give the explicit formula for the operator $\hat\varphi$:
\begin{equation}
(\hat\varphi\psi)(q)=\frac 1{(2\pi h)^n}\int\!\int\varphi(p,(q+y)/2)
e^{i\sum(q_i-y_i)p_i/h}\psi(y)\,dpdy.
\end{equation}
Further, it is not difficult to compute the formula for the {\it $*$-multiplication}
$\varphi_1*\varphi_2$, i.~e., for the function corresponding to composition of operators
$\hat\varphi_1\hat\varphi_2$, so that
$$
(\varphi_1*\varphi_2)\widehat{\ }=\hat\varphi_1\hat\varphi_2.
$$
For that one should compute the product of two operators of kind (27) by formula
(24), and use the inverse Fourier transform (28). Let us give the answer. Denote
\begin{equation}
\{\varphi_1,\varphi_2\}=\sum_{i,j}\omega^{ij}\frac{\partial\varphi_1}{\partial y_i}
\frac{\partial\varphi_2}{\partial y_j},
\end{equation}
where $y_i=q_i$ for $1\le i\le n$ and $y_i=p_{i-n}$ for $n+1\le i\le 2n$, and
$\omega^{ij}=\delta_{i,j-n}-\delta_{i-n,j}$.
Then
\begin{equation}
(\varphi_1*\varphi_2)(y_i)
=\left.\exp\left(-\frac{ih}2\sum_{i,j}\omega^{ij}
\frac{\partial}{\partial y_i}\frac{\partial}{\partial z_j}\right)
\varphi_1(y_i)\varphi_2(z_i)\right|_{z_i=y_i}.
\end{equation}
This product is usually called the {\it Moyal product};  it is not difficult to check
directly that it is associative.

Finally, the same formula (24) implies that the transport equation has the right
form (10) for the Hamiltonian being exponent of a linear form, and hence by linearity
for any Hamiltonian.
Below the Hamiltonian $\hat H$ and other quantum observables will be understood in the
sense of Weyl calculus.

\subsection{Method of complex germ at a point} Following Maslov and Shvedov [2],
let us look for asymptotic solutions of the Schrodinger equation (1) in the form of
wave packets
\begin{equation}
\psi(t,q)=f\left(t,\frac{q-q_0(t)}{\sqrt h}\right)
e^{\frac ih\left(\sum p_0^i(t)(q_i-q_0^i(t))+S(t)\right)}
\end{equation}
for some functions $q_0^i(t)$, $p_0^i(t)$, $S(t)$, $f(t,x)$. Let us call them
{\it Maslov--Shvedov wave packets}.
An example of such wave packet is the Gaussian wave packet (13). We will find equations
on these functions which will imply that the wave packet (32) satisfies the Schrodinger
equation up to $o(h)$.

To this end, note that:

1) action of the operator $\hat q_i-q_0^i$ on the function $\psi$ amounts to action of the
operator $\sqrt hx_i$ on the function $f$;

2) action of the operator $\hat p_i-p_0^i$ on the function $\psi$ amounts to action of the
operator $-i\sqrt h\frac\partial{\partial x_i}$ on the function $f$;

3) the Schrodinger equation (1) for the function $\psi$ amounts, up to $o(h)$, to the
equation
$$
\begin{aligned}{}
&(\sum p_0^i\dot q_0^i-\dot S)f
+\sqrt h\sum(-\dot p_0^ix_i-i\dot q_0^i\frac\partial{\partial x_i})f
+ih\frac{\partial f}{\partial t}\\
&=H(p_0,q_0)f
+\sqrt h\sum(H_{q_i}x_i-iH_{p_i}\frac\partial{\partial x_i})f\\
&+\frac h2\left(\sum H_{q_iq_j}x_ix_j
-iH_{q_ip_j}\left(x_i\frac\partial{\partial x_j}+\frac\partial{\partial x_j}x_i\right)
-H_{p_ip_j}\frac\partial{\partial x_i}\frac\partial{\partial x_j}\right)f
\end{aligned}
$$
(all derivatives of the Hamiltonian are taken at the point $(p_0,q_0)$).
This equation will be satisfied provided the following system of equations holds:
\begin{equation}
\begin{aligned}{}
\sum &p_0^i\dot q_0^i-\dot S=H(p_0,q_0),\\
\dot q_0^i&=H_{p_i},\\
\dot p_0^i&=-H_{q_i},\\
i\frac{\partial f}{\partial t}&=\frac 12\left(\sum H_{q_iq_j}x_ix_j
-iH_{q_ip_j}\left(x_i\frac\partial{\partial x_j}+\frac\partial{\partial x_j}x_i\right)
-H_{p_ip_j}\frac\partial{\partial x_i}\frac\partial{\partial x_j}\right)f.
\end{aligned}
\end{equation}
These equations mean that $(p_0(t),q_0(t))$ is a classical trajectory, $S(t)$ is the action
along this trajectory, and the function $f$ satisfies the Schrodinger equation with the
quadratic Hamiltonian depending on time, with $h=1$.

This latter equation means the following. The quadratic part of the Hamiltonian at each point
of the trajectory gives an infinitesimal symplectic transformation:
\begin{equation}
\begin{aligned}{}
dx_i/dt&=\sum_j H_{q_jp_i}x_j+\sum_j H_{p_ip_j}y_j,\\
dy_i/dt&=-\sum_j H_{q_iq_j}x_j-\sum_j H_{q_ip_j}y_j.
\end{aligned}
\end{equation}
The composition of all these transformations at the time $t$ gives a metaplectic
transformation
\begin{equation}
\left(\left(\begin{array}{cc}
A(t)&B(t)\\
C(t)&D(t)
\end{array}
\right), \sqrt{\det(C(t)Z+D(t))}\right),
\end{equation}
whose action on the function $f(0,x)$ under the Weil representation gives the function $f(t,x)$.

In particular, if we look for the function $f$ in the form of a Gaussian function
\begin{equation}
f(t,x)=\frac1{\sqrt{\det(C(t)Z(0)+D(t))}}\exp\left(\frac i2\sum Z_{ij}(t)x_ix_j\right),
\end{equation}
then for the function
\begin{equation}
Z(t)=(A(t)Z(0)+B(t))(C(t)Z(0)+D(t))^{-1}
\end{equation}
we get a matrix Riccati equation
\begin{equation}
\dot Z+ZH_{pp}Z+H_{pq}Z+ZH_{pq}^T+H_{qq}=0
\end{equation}
of type (21). The matrix $Z(t)$ is called the {\it complex germ}.

It is rather interesting to express these equations in terms of the functions $q(t),q'(t)$
and the Lagrange function $F(t,q,q')$,
i.~e., to rewrite them in the language of the variational principle.
Then equation (34) turns into the Jacobi equation in the theory of second variation,
and equation (38) turns into the corresponding matrix Riccati equation,
see Gelfand--Fomin's book [11], cf. [12].
But in the variational calculus the matrix $Z$ is real; this case will be considered
below.

\subsection{Method of canonical operator}
In conclusion we shall briefly discuss the powerful method of canonical operator,
due to V.~P.~Maslov.
This method allows one, for example, to write out the asymptotic
solution of the Cauchy problem for the Schrodinger equation (see 1.2). Formulas
(2), (8), (3) yield this solution for $t$ sufficiently small,
when different characteristics do not intersect each other and
$\det\left(\frac{\partial q_i(t)}{\partial q_j(0)}\right)\ne 0$.
Method of canonical operator shows what happens with the solution after passing through
{\it focal points}, where characteristics intersect each other and the determinant vanishes.
To this end, let us represent the solution (2) as a superposition of wave packets (32)
satisfying equations (33):
\begin{equation}
\psi(q)=\int e^{\frac ih(S(\alpha)+p_0(\alpha)(q-q_0(\alpha)))}
f\left(\alpha,\frac{q-q_0(\alpha)}{\sqrt h}\right)\frac{d\alpha}{h^{n/2}},
\end{equation}
$\alpha=(\alpha_1,\ldots,\alpha_n)$.

Let us first consider the case when the $n$-dimensional submanifold
$(p_0(\alpha)$, $q_0(\alpha))$
of the phase space $(p,q)$ diffeomorphically projects onto the $q$-plane.
Let us develop the expression under the exponent into the Taylor series
in a vicinity of the point $\alpha_0$ for which $q_0(\alpha_0)=q$, and let us make
change of variables
$$
\frac{\alpha_0-\alpha}{\sqrt h}=y.
$$
We obtain
\begin{equation}
\begin{aligned}{}
&S(\alpha)+p_0(\alpha)(q-q_0(\alpha))=S(\alpha_0)
+\sqrt h\sum_i\left(-\frac{\partial S}{\partial\alpha_i}
+\sum_jp_0^j\frac{\partial q_0^j}{\partial\alpha_i}\right)y_i\\
&+h\sum_{i,i'}\left(\frac12\frac{\partial^2S}{\partial\alpha_i\partial\alpha_{i'}}-
\sum_j\left(\frac{\partial p_0^j}{\partial\alpha_i}\frac{\partial q_0^j}{\partial\alpha_{i'}}
+\frac12p_0^j\frac{\partial^2q_0^j}{\partial\alpha_i\partial\alpha_{i'}}\right)\right)y_iy_{i'}
+O(h^{3/2}).
\end{aligned}
\end{equation}
Hence $\psi(q)$ will not decrease more rapidly than any power of $h$ only in the case when
\begin{equation}
\frac{\partial S}{\partial\alpha_i}=\sum_j p_0^j\frac{\partial q_0^j}{\partial\alpha_i},\ \
1\le i\le n.
\end{equation}
Below we will assume that this equality holds identically. It implies that
\begin{equation}
\sum_j\left(\frac{\partial p_0^j}{\partial\alpha_i}\frac{\partial q_0^j}{\partial\alpha_{i'}}-
\frac{\partial p_0^j}{\partial\alpha_{i'}}\frac{\partial q_0^j}{\partial\alpha_i}\right)=0
\end{equation}
for all $i,i'$. In other words, the symplectic differential 2-form
\begin{equation}
\omega=\sum dp_i\wedge dq_i
\end{equation}
vanishes on the submanifold $(p_0(\alpha),q_0(\alpha))$ of the phase space.
Maslov called such submanifolds Lagrangian. Conversely,
any Lagrangian submanifold diffeomorphically projecting onto the $q$-plane,
is the graph of the differential of some function $S(q)$:
\begin{equation}
p_i(q)=\frac{\partial S}{\partial q_i},\ \ 1\le i\le n.
\end{equation}
Equality (41) also implies that
\begin{equation}
\begin{aligned}{}
&\psi(q)=e^{\frac{iS(\alpha_0)}h}\int e^{-\frac i2\sum\frac{\partial p_0^j}{\partial\alpha_i}
\frac{\partial q_0^j}{\partial\alpha_{i'}}y_iy_{i'}}
f\left(\alpha_0,\sum_i\frac{\partial q_0^j}{\partial\alpha_i}y_i\right)dy\\
&+O(\sqrt h)
=e^{\frac{iS(\alpha_0)}h}\int e^{-\frac i2y^TQ^TPy}f(\alpha_0,Qy)dy+O(\sqrt h)\\
&=\frac{e^{\frac{iS(\alpha_0)}h}}{|\det Q|}
\int e^{-\frac i2x^TPQ^{-1}x}f(\alpha_0,x)dx+O(\sqrt h)\\
&=e^{iS(q)/h}a(q)+O(\sqrt h),
\end{aligned}
\end{equation}
where $P_i^j=\frac{\partial p_0^j}{\partial\alpha_i}$,
$Q_i^j=\frac{\partial q_0^j}{\partial\alpha_i}$.

Let us now assume that the Lagrangian submanifold has been transformed by the Hamiltonian
flow (7) on the phase space at the time $t$.
What then happens with the functions $S(q)=S(\alpha_0)$ and $a(q)$?

Recall that on each trajectory the tangent metaplectic transformation (35) arises.
Denote it by $g$. Then in the formula (45) the following changes will occur:

1) $p_0,q_0$ are transformed by the flow;

2) $S(\alpha_0)\to \widetilde S(\alpha_0)=S(\alpha_0)+$action along the trajectory;

3) $P\to AP+BQ$, $Q\to CP+DQ$;

4) $f(\alpha_0,x)\to\rho(g)f(\alpha_0,x)$.

How is $a(q)$ transformed? To answer this question let us study what is
$\rho(g)e^{\frac i2x^TPQ^{-1}x}$, as promised at the end of 1.5.

Denote $Z=PQ^{-1}$. Now $Z$ is real. Moreover, $\psi_Z(x)=e^{\frac i2x^TZx}$
does not belong now to the Schwartz space and to $L_2$.
It is the unique, up to proportionality, distribution solution of the system of equations
\begin{equation}
(i\frac\partial{\partial x_i}+\sum_j Z_{ij}x_j)\psi_Z=0,\ \ 1\le i\le n.
\end{equation}
The transformation $g$ takes these equations to the equations
\begin{equation}
\sum_j \left(v_{ij}x_j+w_{ij}i\frac{\partial}{\partial x_j}\right)\psi=0,\ \ 1\le i\le n,
\end{equation}
which, for $\det(CZ+D)\ne0$, are equivalent to the equations on the Gaussian function
$\psi_{(AZ+B)(CZ+D)^{-1}}$. Moreover, as it is shown by taking the limit in
equalities (13), (14) as $\Imm Z\to0$, $p_0=0$ ($\Imm Z$ is the imaginary part of the matrix
$Z$), for $\det(CZ+D)\ne0$ we have
\begin{equation}
\begin{aligned}{}
\rho(g)\psi_Z&=\lim_{\Imm Z\to+0}\frac1{\sqrt{\det(CZ+D)}}\psi_{(AZ+B)(CZ+D)^{-1}}=\\
&=\frac{e^{i\pi k/2}}{\sqrt{|\det(CZ+D)|}}\psi_{(AZ+B)(CZ+D)^{-1}}
\end{aligned}
\end{equation}
for some integer $k$ called the {\it Maslov index}. This index has a purely
algebraic definition, see, for example, Hormander's book [10], \S21.6.
As far as we know, the above simple definition of the Maslov index
did not appear in the literature.

In the general case (when $\det(CZ+D)$ can equal 0) it is easy to see that
the system of equations (47) forms a basis of a (real) Lagrangian subspace
$L$ in the $2n$-dimensional vector space with the basis
\begin{equation}
(x_1,\ldots,x_n,i\frac{\partial}{\partial x_1},\ldots,
i\frac{\partial}{\partial x_n}),
\end{equation}
i.~e., an $n$-dimensional subspace on which the symplectic form, given by the
commutator of operators, vanishes. It is not difficult to see that the solution
$\psi(x)=\psi_L(x)$ of this system is, in general case, up to a constant factor,
the product of a function of type $\psi_Z$ of part of the variables, for a
real $Z$, and the delta function of the remaining variables (after an appropriate
linear change of coordinates $x$). The most degenerate case is the system of equations
$x_i\psi=0$, $1\le i\le n$, whose solution is the delta function $\delta(x)$.

Thus, we have described the $\Mp(2n,\R)$-orbit of functions $\psi_Z$ for real $Z$,
or, which is the same, the orbit of the function $1$ in the projectivization $PS'$
of the space $S'$ of distributions. This orbit is isomorphic to the variety of
real Lagrangian subspaces of the $2n$-dimensional symplectic vector space.
This variety is called the {\it Lagrangian Grassmannian}; denote it by $\Lambda_n$.
The embedding $\Lambda_n\to PS'$
induces a complex line bundle $\mu$ on the Grassmannian $\Lambda_n$, whose fiber
at the point $L\in\Lambda_n$ is the line $\CC\psi_L$. This bundle has an action
of the group $\Mp(2n,\R)$. Trivializations and transition functions of this bundle
can be obtained from formula (48).
Let us call this bundle $\mu$ the {\it Maslov bundle} (Hormander [10], \S21.6,
uses another terminology and calls the bundle $\mu$ the tensor product of the half
densities bundle and the Maslov bundle).

Returning to formula (45), we see (assuming that $f(\alpha_0,x)$ belongs to the Schwartz
space with respect to $x$) that under the action of the Hamiltonian flow the function
$a(q)$ is multiplied by
\begin{equation}
\begin{aligned}{}
\frac{|\det Q|\sqrt{|\det(CPQ^{-1}+D)|}}{e^{i\pi k/2}|\det(CP+DQ)|}
&=\frac{e^{-i\pi k/2}}{\sqrt{|\det(CPQ^{-1}+D)|}}\\
&=\frac{e^{-i\pi k/2}}{\sqrt{\left|\det\frac{\partial q_i(t)}{\partial q_j(0)}\right|}},
\end{aligned}
\end{equation}
if after the transformation by the flow the Lagrangian manifold still diffeomorphically
projects onto the $q$-plane (i.~e., if $\det(CP+DQ)\ne0$). This result generalizes
formula (8).

Let us now return to the integral (39) and consider it in the case when the manifold
$(p_0(\alpha)$, $q_0(\alpha))$ not necessarily diffeomorphically projects onto the $q$-plane.
In this case let us present the function $f(\alpha,x)$ as sum of functions $f_l$,
each of which has the support with respect to the variable $\alpha$
diffeomorphically projecting onto some Lagrangian plane in the phase space. Then
let us apply to these functions linear Hamiltonian flows (15),
giving metaplectic transformations $g_l$,
so that these Lagrangian planes get to the $q$-plane. After that let us apply
formula (45). We obtain a wave function $\psi_l(q)$. Finally, let us apply to these
wave functions the transformations $\rho(g_l^{-1})$, and let us take the sum of them.
The obtained wave function $\psi(q)$
(which is, in general, a distribution) is defined correctly up to $O(\sqrt h)$ if
the Maslov index of any closed curve on the Lagrangian manifold is divisible by 4.
The role of function $a(q)$ is played here by a section of the Maslov bundle on
the Lagrangian manifold, induced from the bundle $\mu$ on the Lagrangian Grassmannian.

Thus, the method of canonical operator assigns a (distribution) wave function $\psi(q)$,
defined up to $O(\sqrt h)$, to a Lagrangian submanifold and to a section of the Maslov
bundle on it.
In the case when the Lagrangian submanifold is the graph of
differential of a function $S(q)$,
the wave function has the form (45).
Under the evolution given by the Schrodinger equation, the corresponding Hamiltonian
flow on the phase space transforms the Lagrangian manifold and the section of the
Maslov bundle on it, and hence transforms the wave function $\psi(q)$.
This gives the global asymptotic solution of the Cauchy problem.

The method of canonical operator has far generalizations.
For example, if one integrates not along $n$-dimensional but along $k$-dimensional ($k<n$)
isotropic submanifold, then one obtains the {\it method of canonical operator on a
Lagrangian manifold with complex germ}. The method of complex germ also applies
to approximate solution of linear and even non-linear partial differential equations,
and not only to the Schrodinger equation. See Maslov's books [3--5].

\section{The Weyl algebra in quantum field theory}

\subsection{The Schrodinger equation}
Consider a field theory action functional of the form
\begin{equation}
J=\int_D F(x^0,\ldots,x^n,u^1,\ldots,u^m,u^1_{x^0},\ldots,u^m_{x^n})\,
dx^0\ldots dx^n,
\end{equation}
where $x^0=t,x^1,\ldots,x^n$ are the independent variables, $u^1,\ldots,u^m$ are the dependent
variables, $u^i_{x^j}=\frac{\partial u^i}{\partial x^j}$,
and integration goes over an $(n+1)$-dimensional surface $D$
(the graph of the functions $u^i(x)$)
with the boundary $\partial D$ in the space $\R^{m+n+1}$.
The main simplest example for our considerations is the $\varphi^4$ model in four
dimensions:
\begin{equation}
J=\int\left(\frac12\left(u_t^2-\sum_{j=1}^3 u_{x^j}^2-m^2u^2\right)
-\frac1{4!}gu^4\right)\,dtdx^1dx^2dx^3.
\end{equation}
The Schrodinger equation for the model (51) reads
\begin{equation}
ih\frac{\partial\Psi}{\partial t}=\int\widehat H\left(t,\x,u^i(\x),
\frac{\partial u^i}{\partial\x},-ih\frac{\delta}{\delta u^i(\x)}\right)
\Psi\,d\x.
\end{equation}
Here $\x=(x_1,\ldots,x_n)$;

$\Psi$ is an unknown complex valued functional of the variable
$t$ and of functions $u^i(\x)$, $1\le i\le m$;

$H$ is the density of the Hamiltonian of the theory, which equals the
Legendre transform of the Lagrangian $F$ with respect to the variables $u^i_t$; denote
the dual variables to $u^i_t$ by $p^i$;

$\widehat H$ is the density of the quantum Hamiltonian,
obtained from $H$ by the substitution of the variational differentiation operator
$-ih\frac{\delta}{\delta u^i(\x)}$ instead of $p^i$.

As in quantum mechanics, here the problem arises of ordering of the operators
$u^i(\x)$ and $-ih\frac{\delta}{\delta u^i(\x)}$ in the quantum Hamiltonian. Let us not
consider this problem now, all the more in the case of the $\varphi^4$ model there is
no such ambiguity.

The Schrodinger equation (53) is obviously relativistically non-in\-var\-iant. But one
can write out a relativistically invariant version of this equation, in which
the surface $t=\const$ in the space-time is changed by an arbitrary space-like
surface, and the functional $\Psi$ depends on this surface and on functions
$u^i(s)$ on it, where $s=(s_1$, $\ldots$, $s_n)$ are parameters on the surface.
This relativistically invariant version is obtained in exactly the same way as the usual
quantum mechanical Schrodinger equation, by the formal substitution into the
generalized field theory Hamilton--Jacobi equation, see [6]. In physical literature
a close equation is called the Tomonaga--Schwinger equation [13], and the problem of
solving this equation is called quantization on space-like surfaces.

One can give a rigorous mathematical sense to the Schrodinger equation (53)
and its relativistically invariant generalization. To this end, the conventional usual way
is to consider weakly continuously differentiable sufficient number of times
functionals on a nuclear space of functions $u^i(s)$, for example, on the Schwartz
space. In this interpretation, for example, the variational derivative
$\frac{\delta\Psi}{\delta u^i(\x)}$ is a distribution in $\x$,
the second variational derivative $\frac{\delta^2\Psi}{\delta u^i(\x)\delta u^{i'}(\x')}$
is a distribution in $(\x,\x')$, etc.

However, it is not difficult to see that with such understanding the Schrodinger equation,
say, for the $\varphi^4$ model,
\begin{equation}
ih\frac{\partial\Psi}{\partial t}
=\int\left(-\frac{h^2}2\frac{\delta^2}{\delta u(\x)^2}+\frac12(\grad u(\x))^2
+\frac{m^2}2u(\x)^2+\frac g{4!}u(\x)^4\right)\Psi\,d\x,
\end{equation}
does not have nonzero four times differentiable solutions. Indeed, consider the
second derivative $\frac{\partial^2\Psi}{\partial t^2}$.
In the expression for this derivative following from equation (54),
we will have the term
$$
\int\!\int\frac{\delta^2}{\delta u(\x)^2}u(\y)^2\Psi\,d\x d\y,
$$
which, as it is easy to see, has no sense (the second variational derivative
cannot be restricted as a distribution to the diagonal $\x=\x'$).

One can give physical arguments as well in favor of the statement that states cannot
be functionals. Indeed, if it were so (as it was assumed in past, see, for example,
[14,15], etc.), then the values of these functionals or related quantities, in principle,
could be measured. On the other hand, it is known (see, for example,
\S1 of the book [16] by Berestetsky, Lifschitz, and Pitaevsky) that in relativistic
quantum dynamics, quantum mechanical quantities like energy and momentum are
theoretically non-measurable, and the only measurable quantities are the scattering sections.

Besides that, one would like to have that in the case of free scalar field given by a
quadratic Hamiltonian, the Schrodinger equation be solved exactly, similarly to
the finite dimensional case. This implies that the space of states and the space of
operators have an action of infinite dimensional symplectic group. Indeed,
the evolution operators of classical field equations from one space-like surface
to another are canonical transformations, preserving the field theory Poisson bracket.
This follows from the generalized field theory canonical Hamilton equations
(see [6]).
In the case of free field these operators are linear, i.~e., symplectic. Hence
the action of quantum Hamiltonians should admit a compatible action of a group
of symplectic transformations of the space of functions $(u^i(s)$, $p^i(s))$.

The traditional action of an infinite dimensional symplectic group is the
projective Segal--Shale--Weil--Berezin representation in the Fock space
[14,15]. However, this infinite dimensional symplectic group does not suit for
our purposes, as shown in the important paper [17]. In this paper it is shown
that the evolution operators of the Klein--Gordon equation from one
space-like surface to another, in general, do not belong to that version of
infinite dimensional symplectic group which acts on the Fock space.
This is also in accordance with the physical arguments above.
The evolution operators belong to the group of continuous symplectic transformations
of the Schwartz space of functions
$(u^i(s),p^i(s))$. It is this group that should act on the space of quantum Hamiltonians.

To achieve this, it is natural, as in \S1, to introduce, instead of the algebra of
differential operators on the space of functions, the infinite dimensional
generalization of the Weyl algebra. Let us give a definition of this
generalization.

\subsection{Infinite dimensional Weyl algebra}
\subsubsection{Definition of the Weyl algebra}

The Weyl algebra is constructed starting from a symplectic vector space.
Consider the symplectic Schwartz space of rapidly decreasing functions
$(u^i(s),p^i(s))$ with the Poisson bracket
\begin{equation}
\{\Phi_1,\Phi_2\}=\sum_i\int\left(\frac{\delta \Phi_1}{\delta u^i(s)}
\frac{\delta \Phi_2}{\delta p^i(s)}-\frac{\delta \Phi_1}{\delta p^i(s)}
\frac{\delta \Phi_2}{\delta u^i(s)}\right)ds
\end{equation}
of two functionals $\Phi_l(u^i(\cdot),p^i(\cdot))$, $l=1,2$.
Let us write it in the form
\begin{equation}
\{\Phi_1,\Phi_2\}=\int\sum_{i,j}\omega^{ij}\frac{\delta \Phi_1}{\delta y^i(s)}
\frac{\delta \Phi_2}{\delta y^j(s)}\,ds,
\end{equation}
where $y^i=u^i$ for $1\le i\le m$ and $y^i=p^{i-m}$ for $m+1\le i\le 2m$, and
$\omega^{ij}=\delta_{i,j-m}-\delta_{i-m,j}$, as in 1.4.
The Weyl algebra is defined as the algebra of infinitely differentiable functionals
$\Phi(u^i(\cdot),p^i(\cdot))$
with respect to the Moyal $*$-product
\begin{equation}
\begin{aligned}{}
&(\Phi_1*\Phi_2)(y^i(\cdot))\\
&=\left.\exp\left(-\frac{ih}2\int\sum_{i,j}\omega^{ij}
\frac{\delta}{\delta y^i(s)}\frac{\delta}{\delta z^j(s)}\,ds\right)
\Phi_1(y^i(\cdot))\Phi_2(z^i(\cdot))\right|_{z^i(\cdot)=y^i(\cdot)}.
\end{aligned}
\end{equation}
This product is not everywhere defined: for example, $u^i(s)*p^i(s)$ is undefined.
We shall not go into details of the domain of multiplication, as well as into
details of defining topology in the Weyl algebra.
This should be the subject of a separate investigation. Note only that
if all necessary integrals and series are defined and absolutely convergent,
then the $*$-product is associative.
This is a formal check similar to the finite dimensional case.
In this paper we will be interested only in some concrete
computations in the Weyl algebra. In algebraic quantum field theory
[17,18,19] a somewhat different definition of Weyl algebra is adopted.

Below we will see that the Weyl algebra allows one to construct a logically
self-consistent theory of free quantum scalar field and to simplify drastically
perturbative theory of interacting quantum fields.

\subsubsection{The problem of states}

Thus, operators in equations (53), (54), and others will be understood as elements of
the Weyl algebra. And how will be understood states $\Psi$?
They already cannot be functionals of $u^i(\cdot)$, since
the Weyl algebra does not act on them.
In the case of a finite dimensional symplectic vector space,
the Weyl algebra acts canonically on {\it half-forms} on a Lagrangian subspace.
In coordinates $q_1,\ldots,q_N,p_1,\ldots,p_N$ half-forms look $f(q_1$, $\ldots$, $q_N)$
$(dq_1\ldots dq_N)^{1/2}$.
The fact that the Weyl algebra acts on half-forms, can be seen, for example, as follows.
The operator $\frac ih(p_iq_j)\widehat{\ }$ of infinitesimal linear change of
coordinates from the Weyl algebra acts as
$$
\frac12\left(q_j\frac{\partial}{\partial q_i}+\frac{\partial}{\partial q_i}q_j\right)
=q_j\frac{\partial}{\partial q_i}+\frac12\delta_{ij},
$$
and this is the action on half-forms.

What are half-forms on an infinite dimensional space of functions $u^i(s)$?
Seemingly, one cannot say anything
definite at this point. At least, half-forms cannot be constructed from
finite dimensional spaces,
analogously to the construction of measures on an infinite dimensional space.
Author's attempts to construct half-forms failed to be success (see, for example, [20]).

But actually, in order to obtain physically important quantities for free field,
we need not states:
it suffices to use only operators, as will be shown below.
States are ``non-observable neither physically
nor mathematically''. Hence we will consider, instead of equations (53), (54), the
Heisenberg equation
for an element $\Phi(t;u^i(\cdot),p^i(\cdot))$ of the Weyl algebra:
\begin{equation}
ih\frac{\partial\Phi}{\partial t}=\left[\int\!H(t,\x,u^i(\x),
\frac{\partial u^i}{\partial\x},p^i(\x))d\x,\Phi\right]
\end{equation}
and its relativistically invariant generalization, where
\begin{equation}
[\Phi_1,\Phi_2]=\Phi_1*\Phi_2-\Phi_2*\Phi_1
\end{equation}
is the commutator in the Weyl algebra. The classical limits of equations (58) are
the field theory Hamilton equations
\begin{equation}
\frac{\partial\Phi}{\partial t}=\{\Phi,\int\!H\,d\x\},
\end{equation}
equivalent to the Euler--Lagrange equations.

\subsection{Quantization of free scalar field}

\subsubsection{Solution of the Heisenberg equation for free scalar field}

Solution of equation (58) is given by the formal equality
\begin{equation}
\Phi(t_1)=U(t_0,t_1)*\Phi(t_0)*U(t_0,t_1)^{-1},
\end{equation}
where
\begin{equation}
U(t_0,t_1)=T\exp\int_{t_0}^{t_1}\!\int\frac1{ih}H(t,\x)\,dtd\x,
\end{equation}
and $T\exp\int$ means the ordered exponent (the multiplicative integral):
\begin{equation}
T\exp\int_{t_0}^{t_1}\Gamma(t)dt
=1+\int\limits_{t_0<t<t_1}\Gamma(t)dt+\int\limits_{t_0<t'<t<t_1}\Gamma(t)*\Gamma(t')\,dtdt'
+\ldots.
\end{equation}
(Cf. 2.4.1 below.)

Let us first consider the free scalar field ($g=0$). In this case
$$
H(t,\x)=\frac12(p(\x)^2+(\grad u(\x))^2+m^2u(\x)^2)
$$
is a quadratic expression not depending on $t$,
hence we can omit the sign $T$ before exponent.
Due to the fact that the Hamiltonian
\begin{equation}
H_0=\int H(\x)d\x
\end{equation}
is quadratic, we have
\begin{equation}
\frac1{ih}[H_0,\Phi]=\{\Phi,H_0\},
\end{equation}
therefore,
$\Phi(t_1; u(\cdot),p(\cdot))$ is obtained from
$\Phi(t_0;u(\cdot),p(\cdot))$ by the linear symplectic change of
variables
\begin{equation}
(u(t_0,\x),p(t_0,\x)=u_t(t_0,\x))\to(u(t_1,\x),p(t_1,\x)=u_t(t_1,\x))
\end{equation}
given by the evolution operator of the canonical Hamilton equations,
i.~e., by the evolution operator of the Klein--Gordon equation
\begin{equation}
\Box u-m^2u=-\frac{\partial^2u}{\partial t^2}+\sum_{j=1}^n\frac{\partial^2u}{(\partial x^j)^2}
-m^2u=0
\end{equation}
from the Cauchy surface $t=t_0$ to the Cauchy surface $t=t_1$.
Here it is rather essential that the evolution operator
is a continuous linear invertible operator in the Schwartz space of functions
$(u(\x)$, $p(\x))$.
Similar statement is true for the evolution of the functional
$\Phi$ between any two space-like Cauchy surfaces.
(For non-quadratic Hamiltonians and non-linear classical evolution operators
similar statement is not true.)

Hence we can identify the Weyl algebras corresponding to different space-like surfaces,
by means of the evolution operators of the Klein--Gordon equation. (Cf. [17].)
In other words, we can consider the Weyl algebra $W_0$
of the symplectic vector space of solutions $u(t,\x)$
of the Klein--Gordon equation on the whole space-time.
The symplectic form on this vector space is given by taking the Cauchy data
\begin{equation}
u(t,\x)\to(u(s),p(s))
\end{equation}
on any space-like surface $x=x(s)$.
(The quantity $p(s)$ is proportional to the normal derivative of the function $u(t,\x)$
at the point $s$.)
Below we will fix this identification of the Weyl algebras of various space-like surfaces.

\subsubsection{Green functions}

Let us now consider the free scalar field with a source, i.~e. put
\begin{equation}
H(t,\x,u,p)=\frac12(p(\x)^2+(\grad u(\x))^2)+\frac{m^2}2u(\x)^2+\jj(t,\x)u(\x),
\end{equation}
where $\jj(t,\x)$ is a smooth function with compact support (a source).
Denote the corresponding formal element (62) of the Weyl algebra by $U_\jj(t_0,t_1)$,
and the Hamiltonian by
\begin{equation}
H_\jj(t)=H_0+\int \jj(t,\x)u(\x)\,d\x,
\end{equation}
to show dependence on the source.
Then, if the support of the function $\jj(t,\x)$ is situated between the planes
$t=t_{\min}$ and $t=t_{\max}$, then
the formal element
\begin{equation}
R_\jj(t_0)=U_0(t_{\max},t_0)*U_\jj(t_{\min},t_{\max})*U_0(t_0,t_{\min})
\end{equation}
of the Weyl algebra does not depend on $t_{\min},t_{\max}$. Besides that, we have
\begin{equation}
R_\jj(t_1)=U_0(t_0,t_1)*R_\jj(t_0)*U_0(t_0,t_1)^{-1}.
\end{equation}
Hence the element $R_\jj(t_0)=R_\jj(t_0;u(\cdot),p(\cdot))$ correctly defines an element
of the Weyl algebra of any space-like surface under our identification, i.~e.
an element $R_\jj$ of the Weyl algebra $W_0$.
This element equals
\begin{equation}
R_\jj=R_\jj(u(\cdot,\cdot))
=T\exp\int\limits_{-\infty}^{\infty}\!\int\frac1{ih}\jj(t,\x)u(t,\x)\,dtd\x,
\end{equation}
where
\begin{equation}
u(t,\x)=\exp\left(-\frac{t-t_0}{ih}H_0\right)*u(\x)*\exp\left(\frac{t-t_0}{ih}H_0\right)
\end{equation}
is understood as a functional on the space of solutions $u(\cdot,\cdot)$
of the Klein--Gordon equation, i.~e. as an element of the algebra
$W_0$. (See 2.4.1 below.)
Let us emphasize that expression (74) is purely symbolic, since the element
$\exp(t_1-t_0)H_0/(ih)$ does not exist in the Weyl algebra, because already
$H_0*H_0$ does not exist.
Let us call the element $R_\jj$ the {\it generating functional of operator
Green functions of a free field}.
Let us also call the coefficients of the Taylor decomposition of the functional
$R_\jj$ with respect to $\jj$ at the point $\jj\equiv0$,
\begin{equation}
(ih)^N\left.\frac{\delta^NR_\jj}{\delta\jj(t_1,\x_1)\ldots\delta\jj(t_N,\x_N)}
\right|_{\jj\equiv0}
=Tu(t_1,\x_1)*\ldots*u(t_N,\x_N),
\end{equation}
by the operator Green functions of a free field; here
the symbol $T$ denotes $*$-product ordered by decreasing of the variables $t_i$.
The operator Green functions are distributions of $(t_1,\x_1)$, $\ldots$, $(t_N,\x_N)$
with values in $W_0$, symmetric with respect to permutations of indices.

Let us now pass to the scalar Green functions. To this end, define a linear functional
on the algebra $W_0$, called the vacuum average of an element $\Phi$ from $W_0$
and denoted by $\langle\Phi\rangle$ or $\langle0|\Phi|0\rangle$,
in the following way. The momentum representation
\begin{equation}
\tilde u(p_0,\ldots,p_n)=\frac1{(2\pi)^{(n+1)/2}}\int e^{-i\sum p_jx^j}u(x^0,\ldots,x^n)\,dx
\end{equation}
of a solution $u(t,\x)$ of the Klein--Gordon equation, where $t=x^0$, is a distribution
supported on two sheets of the mass surface $p_0=\pm\sqrt{\p^2+m^2}$, where
$\p=(p_1,\ldots,p_n)$;
in this paper we restrict ourselves by theories with {\it nonzero mass}, $m>0$. Hence
$u(t,\x)$ can be uniquely decomposed into the sum
\begin{equation}
u(t,\x)=u_+(t,\x)+u_-(t,\x)
\end{equation}
of a positive frequency solution $u_+(t,\x)$, whose Fourier transform is
supported on the sheet $p_0>0$,
and a negative frequency solution $u_-(t,\x)$, whose Fourier transform is supported
on the sheet $p_0<0$.
We have
\begin{equation}
\begin{aligned}{}
&[u_+(t_1,\x_1),u_+(t_2,\x_2)]=[u_-(t_1,\x_1),u_-(t_2,\x_2)]=0,\\
&[\tilde u_-(p),\tilde u_+(p')]=-h\delta(p+p')\delta(p^2-m^2),\ \ p_0<0,\ \ p_0'>0,
\end{aligned}
\end{equation}
where $p^2=p_0^2-\sum_{j=1}^np_j^2$.
Define $\langle\Phi\rangle$ as the unique (not everywhere defined) functional with the
following properties:
\begin{equation}
\langle\Phi*u_-(t,\x)\rangle=\langle u_+(t,\x)*\Phi\rangle=0,\ \ \langle1\rangle=1.
\end{equation}
Define the Green functions by the equality
\begin{equation}
\langle u(t_1,\x_1)\ldots u(t_N,\x_N)\rangle=\langle Tu(t_1,\x_1)*\ldots*u(t_N,\x_N)\rangle,
\end{equation}
and their generating functional by the equality
\begin{equation}
Z(\jj)=\langle R_\jj\rangle.
\end{equation}
A computation left to the reader (apply Fourier transform with respect to $\x$; cf.
the textbook [21] by Bogolyubov and Shirkov) shows that the two-point Green function
turns out to be equal to the Feynman propagator
\begin{equation}
\langle u(t,\x)u(t',\x')\rangle\ \widetilde{ }\ =ih\frac{\delta(p+p')}{p^2-m^2+i\varepsilon},
\end{equation}
and the generating functional of the Green functions is given by the usual expression
\begin{equation}
Z(\jj)=\exp\frac{-i}{2h}\int\frac{\tilde\jj(p)\tilde\jj(-p)}{p^2-m^2+i\varepsilon}\,dp.
\end{equation}

\subsubsection{The Fock space}

Define the standard {\it Fock space}, linearly generated by the vectors
\begin{equation}
|p_{(1)},\ldots,p_{(N)}\rangle=\tilde u_+(p_{(1)})\ldots\tilde u_+(p_{(N)})|0\rangle
\end{equation}
(after integration over $p_{(i)}$ with a function $f(p_{(1)}$, $\ldots$, $p_{(N)})$)
for all $N$, $p_{(1)}$, $\ldots$, $p_{(N)}$ such that $p_{(i)}^2=m^2$,
$p_{0(i)}>0$.
In other words, the Fock space is the direct sum over all $N$ of spaces of symmetric
functions of $N$ variables $p_{(i)}$. On this space one introduces the structure of
a Hilbert space, namely, the direct sum over all $N$ of the spaces $L_2$ of symmetric
functions of $N$ variables $p_{(i)}$
with respect to the natural Lorentz-invariant measure
\begin{equation}
\delta(p^2-m^2)dp=\frac{d\p}{2p_0}
\end{equation}
on the mass surface $p^2=m^2$.

One can formally assign an operator in the Fock space to an element $\Phi$ of the Weyl algebra
$W_0$, with the matrix elements
\begin{equation}
\langle0|\tilde u_-(-p'_{(1)})\ldots\tilde u_-(-p'_{(N')})*\Phi*\tilde u_+(p_{(1)})\ldots
\tilde u_+(p_{(N)})|0\rangle.
\end{equation}
But for many important operators responsible for local dynamics, for example,
for the Hamiltonian $\Phi=H_0$, the expression (86) is undefined.

Note two properties of this correspondence, which it is not difficult to check.

1) $*$-product of functionals corresponds to composition of operators, so that this
correspondence is a (not everywhere defined)
homomorphism of the Weyl algebra $W_0$ to the algebra of operators in the Fock space.

2) Complex conjugation of functionals goes to the Hermitian conjugation of operators in
the Hilbert space. In particular, the operator $\tilde u_+(p)$ is Hermitian conjugate
to the operator $\overline{\tilde u_+(p)}=\tilde u_-(-p)$.

\subsection{Quantization of interacting fields}

We start with the formal decomposition of a solution of the Heisenberg equation (58)
in the $\varphi^4$ model into the perturbation series with respect to the coupling constant $g$. For that, recall the perturbation theory of linear differential equations.

\subsubsection{Perturbation theory of linear differential equations}

Consider the equation
\begin{equation}
\frac{dv}{dt}=A(t)v(t)+B(t)v(t),
\end{equation}
where $A(t)$ is a linear operator applied to a vector $v(t)$, and $B(t)$ is a possibly
nonlinear operator which is considered as a small perturbation. Let
\begin{equation}
U_0(t_1,t_2)=T\exp\int_{t_1}^{t_2}A(t)\,dt
\end{equation}
be the evolution operator of the non-perturbed equation. Let us find the series
for the evolution operator $U(t_1,t_2)$
of the perturbed equation (87) from time $t_1$ to time $t_2$
by powers of the perturbation $B$. To this end, let us use the formula
\begin{equation}
U(t_1,t_2)=U_0(t_1,t_2)+\int_{t_1}^{t_2}U_0(t,t_2)B(t)U(t_1,t)\,dt.
\end{equation}
Iterating this formula, we shall find a decomposition of the operator $U$,
in the general case, as a sum over trees,
on whose vertices the terms of the Taylor series of the operator $B(t)$ stand, and on
the edges the operators $U_0$ stand.
In the particular case when the operator $B(t)$ is linear we obtain the formula
\begin{equation}
U(t_1,t_2)=U_0(t_0,t_2)\left(T\exp\int_{t_1}^{t_2}\widetilde B(t)\,dt\right)U_0(t_0,t_1)^{-1},
\end{equation}
where
\begin{equation}
\widetilde B(t)=U_0(t_0,t)^{-1}B(t)U_0(t_0,t).
\end{equation}
We have already used this formula in the derivation of the relation (73).

\subsubsection{Formal perturbation series for the Heisenberg equation in the $\varphi^4$ model}

Let us apply this theory for the Heisenberg equation (58) in the $\varphi^4$ model.
Denote the evolution operator of the Heisenberg equation for free field from time $t_0$
to time $t_1$ by $V_0(t_0,t_1)$, so that formally we have
\begin{equation}
V_0(t_0,t_1)\Phi=U_0(t_0,t_1)*\Phi*U_0(t_0,t_1)^{-1},
\end{equation}
where
\begin{equation}
U_0(t_0,t_1)=\exp\frac{t_1-t_0}{ih}H_0.
\end{equation}
(Recall that this formal expression does not exist in the Weyl algebra.) Then,
by (90), the perturbation series for the evolution operator $V(t_1,t_2)$ of the Heisenberg
equation in the $\varphi^4$ model is given by the formula
\begin{equation}
V(t_1,t_2)\Phi=V_0(t_0,t_2)[P(t_1,t_2)*V_0(t_0,t_1)^{-1}\Phi*P(t_1,t_2)^{-1}],
\end{equation}
where
\begin{equation}
\begin{aligned}{}
P(t_1,t_2)&=U_0(t_0,t_2)^{-1}*U(t_1,t_2)*U_0(t_0,t_1)\\
&=T\exp\int_{t_1}^{t_2}\!\int\frac1{ih}gu(t,\x)^4/4!\,dtd\x.
\end{aligned}
\end{equation}
The coefficient before $g^N$ of the latter series equals
\begin{equation}
\int\frac1{(ih)^N4!^N N!} Tu(t_{(1)},\x_{(1)})^4*\ldots*u(t_{(N)},\x_{(N)})^4
\prod dt_{(i)}d\x_{(i)},
\end{equation}
where integration goes over the strip $t_1\le t_{(i)}\le t_2$.
Absolutely the same integral describes the perturbation series for the evolution operator
between any two space-like surfaces, but the integration goes over
the strip between these surfaces.

\subsubsection{Feynman diagrams}
Let us compute the expression under the integral in the Weyl algebra.
To this end, one must firstly find the formula for the product of $N$ elements
of the Weyl algebra. This is left to the reader, starting from the case $N=3$.
Let us give the answer for the expression under the integral (96). It equals
the sum over the {\it Feynman diagrams}, i.~e.
over the 4-valent non-oriented graphs with $N$ vertices,
and to each graph one assigns an element of the Weyl algebra according to the following
rules:

1) to each vertex one assigns the factor $(ih)^{-1}$;

2) to each external tail (i.~e. to an edge with one vertex) one assigns the factor
$u(t_{(i)}$, $\x_{(i)})$, where $i$ is the number of the vertex;

3) to each edge with two vertices $i$ and $j$ one assigns the factor
\begin{equation}
-\frac{ih}2T\{u(t_{(i)},\x_{(i)}),u(t_{(j)},\x_{(j)})\}=-ihD(x_{(i)}-x_{(j)}),
\end{equation}
where $D(x)$ is certain Green function of the Klein--Gordon equation, whose Fourier
transform equals
\begin{equation}
\widetilde D(p)=\PV\frac1{p^2-m^2}
\end{equation}
($\PV$ is the Cauchy principal value);

4) to the whole diagram one assigns the factor $1/M$, where $M$ is the number
of symmetries of the diagram, i.~e. permutations of the vertices and the edges
of the diagram preserving the graph.

After that all factors are multiplied (in the usual sense, and not in the sense of
$*$-product).

Thus, the power of the number $h$ for a Feynman diagram equals the difference
between the number of internal edges and the number of vertices, i.~e. it equals to
the number of independent loops in the diagram minus the number of its connected
components.

We see that, for example, in the case of multiple edges the expression under the
integral contains the square of the function $D(x)$. This is a distribution with
singularities on the light cone, and its square is non-integrable, say, for $n=3$,
because integral of the square of the expression (98)
diverges at large momenta.
Hence the perturbation series is given, in general, by divergent integrals.

But in the {\it tree approximation} (sum over diagrams without loops) we formally
obtain, from the perturbation series for the Heisenberg equation, the
perturbation series for the evolution operator of the non-linear
classical field equation
\begin{equation}
\Box u-m^2u=gu^3/3!.
\end{equation}
The check of this statement is left to the reader as a useful exercise in
perturbation theory.

\subsubsection{An attempt to define dynamical evolution in quantum field theory}

The next attempt to ``quantize fields'' could be an attempt to construct the dynamical
evolution in quantum field theory using, for each space-like surface, some
non-commutative deformation of the algebra of functionals on the phase space
with the Poisson bracket.
Similarly to the linear case, in which we have chosen the deformation being the
Weyl algebra, which admits the symplectic group of transformations, the
required deformation in the general case could be
``adapted'' to the non-linear canonical transformations of the phase space,
given by the evolution operators of the Hamilton equations, i.~e., of the field equations.
To each pair of space-like surfaces one would assign an isomorphism of the
corresponding deformed algebras of functionals, whose classical limit as
$h\to0$ would coincide with the isomorphism of the Poisson algebras of functionals,
given by the transform of the classical evolution.

A possible example of such deformation in the finite dimensional case is the so called
Fedosov deformation quantization of symplectic manifolds [22].
This construction uses the bundle of Weyl algebras on the phase space with the flat
connection ({\it Abelian connection}
in Fedosov's terminology), originating from a symplectic connection on the tangent
bundle to the phase space. However, author's attempts to use this construction
failed to be success. Besides that, in the finite dimensional case any Fedosov
deformation is non-canonically isomorphic to the Weyl algebra.
So it seems that for the purposes of quantum field theory, the Weyl algebra is the most
appropriate deformation of the algebra of functions on the phase space, even in the
non-linear case.

Therefore the next attempt to construct quantization of fields will be an attempt
to construct, for each pair of parameterized space-like surfaces
$\calC_1$, $\calC_2$, an isomorphism of the corresponding Weyl algebras
$W_{\calC_1}\to W_{\calC_2}$. If $\calC_1$, $\calC_2$
are two parameterizations of one and the same space-like surface,
then this isomorphism should coincide with the action of the change of variables
$s$ on functions $u^i(s)$,
$p^i(s)$. (Here $u^i(s)$ are transformed like functions, and
$p^i(s)$ like densities.)
For three space-like surfaces $\calC_1$, $\calC_2$, $\calC_3$
the isomorphism $W_{\calC_1}\to W_{\calC_3}$ should coincide with the composition
of isomorphisms
$W_{\calC_1}\to W_{\calC_2}$ and $W_{\calC_2}\to W_{\calC_3}$. The family of
isomorphisms should be also symmetric with respect to the symmetry group of the theory
(in the case of $\varphi^4$ model this is the Poincare group, i.~e. the group
containing the Lorentz transformations and the parallel translations).
Finally, the classical limit of the isomorphism $W_{\calC_1}\to W_{\calC_2}$
as $h\to 0$ should coincide with the isomorphism of Poisson algebras
of functions given by the classical evolution.

\subsubsection{Dynamical evolution and perturbation theory. The subtraction program}

Let us try to construct the dynamical evolution, as described in the previous subsection,
for the model $\varphi^4$, $n=3$, in the framework of perturbation theory.
Here the main idea should be the physical idea, due to Bethe, exposed at the
beginning of the Introduction to Bogolyubov--Shirkov's book [13], the idea which lead
to the renormalization program. Let us recall this idea in our context. Assume that
the required dynamical evolution exists and describes real physical processes
for interacting fields. But in the framework of perturbation theory,
we can obtain only approximations of some order with respect to the coupling constant,
which by themselves can give divergent quantities, because the field by itself,
without interaction, has no physical sense. And the quantities which do have physical sense,
such as the assumed dynamical evolution, can be given in perturbation theory by
divergent expressions. For example, this means that the quantum Hamiltonian of the
``right'' dynamical evolution equals the classical Hamiltonian plus corrections in
perturbation theory, which can be infinite. The formal purpose, however, is to construct
with the help of these heuristic constructions a ``real'' family of isomorphisms
of Weyl algebras, as pointed out in the preceding subsection.

Thus, the main idea will be an attempt to ``subtract infinities from the perturbation series'',
so as to obtain convergent integrals and so that this subtraction of infinities
have the heuristic sense of adding infinite summands to the quantum Hamiltonian,
which yields a family of (finite) isomorphisms of the Weyl algebras. Note that if we restrict
ourselves by the space-like surfaces $t=\const$, then we just look for a one-parametric
group of automorphisms of the Weyl algebra of functionals
$\Phi(u^i(\x)$, $p^i(\x))$.

\subsubsection{Diagram rules in the $p$-representation}

In order to subtract infinities from the integrals corresponding to Feynman diagrams,
it is convenient first to pass to the momentum representation. Let us state
the rules of writing integrals in the $p$-representation.

1) To each internal edge one assigns some orientation and a 4-momentum $p$, after which
one assigns the factor $-ih\widetilde D(p)$ (98).

2) To each external edge one assigns the orientation from the vertex outside and a 4-momentum
$p$, after which one assigns the factor $\tilde u(p)$.

3) To each vertex one assigns the factor
\begin{equation}
(ih)^{-1}\widetilde\chi(\pm p_{(1)}\pm p_{(2)}\pm p_{(3)}\pm p_{(4)}),
\end{equation}
where $\pm p_{(i)}$ is the momentum outgoing from the vertex along the $i$-th edge
(the sign plus is taken if the edge is oriented outside of the vertex, and the sign minus
in the opposite case); $\widetilde\chi(p)$ is the Fourier transform
of the characteristic function $\chi(x)$ of the strip between the space-like
surfaces, i.~e. $\chi(x)=1$ if $x$ belongs to the strip and $\chi(x)=0$ otherwise.

4) To the whole diagram one assigns the symmetry factor $1/M$.

After that all the factors are multiplied, and one integrates over all the momenta $p$.

\subsubsection{The ``fish'' diagram}
Let us first consider the simplest one-loop diagram ``fish'' with two vertices (Fig.~1).
\begin{figure}[h]\centering
\includegraphics{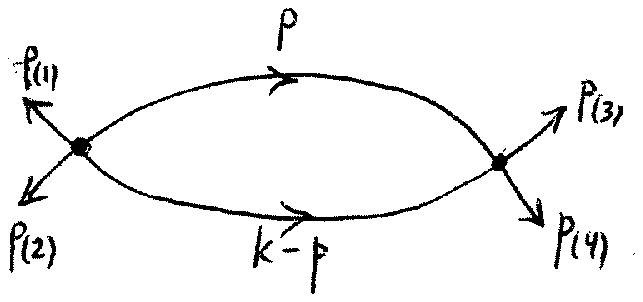}
\caption{}
\end{figure}
To it the following integral corresponds:
\begin{equation}
\int\frac{\widetilde\chi(p_{(1)}+p_{(2)}+k)\widetilde\chi(p_{(3)}+p_{(4)}-k)}
{(p^2-m^2)((k-p)^2-m^2)}\prod_{i=1}^4\tilde u(p_{(i)})dp_{(i)}dpdk,
\end{equation}
in which we have omitted for shortness the constant and the signs $\PV$.
This integral logarithmically diverges for large $p$. The divergence for large $k$
is absent because of the oscillating behavior of the numerator of the fraction.
The divergence with respect to $p_{(i)}$ is also absent because the distribution
$\tilde u(p_{(i)})$ is supported on the mass surface $p_{(i)}^2=m^2$ and rapidly
decreases at infinity, since the function $u(t,\x)$ rapidly decreases at infinity
in space directions.

Note that the function $\widetilde\chi(p)$ satisfies the identity
\begin{equation}
\int\widetilde\chi(q_{(1)}+k)\widetilde\chi(q_{(2)}-k)dk=\widetilde\chi(q_{(1)}+q_{(2)}),
\end{equation}
which is obtained by Fourier transform from the equality $\chi^2=\chi$. If we subtract
from the fraction under the integral the fraction
\begin{equation}
\frac{\widetilde\chi(p_{(1)}+p_{(2)}+k)\widetilde\chi(p_{(3)}+p_{(4)}-k)}
{(p^2-m^2)^2},
\end{equation}
then the integral becomes convergent. Heuristically, from the initial integral we thus
subtract the infinite expression
\begin{equation}
\int\frac{dp}{(p^2-m^2)^2}\int\widetilde\chi(p_{(1)}+p_{(2)}+p_{(3)}+p_{(4)})
\prod_{i=1}^4\tilde u(p_{(i)})dp_{(i)},
\end{equation}
which corresponds to subtraction from the Hamiltonian of the infinite term
\begin{equation}
ihg^2\int\frac{dp}{(p^2-m^2)^2}\cdot\frac{u^4}{4!}.
\end{equation}
Hence at this level the subtraction program gives a correctly defined family of
isomorphisms of the Weyl algebras, satisfying all the necessary requirements.
This family of isomorphisms, however, is defined not uniquely, but only
up to adding a finite summand $ihg^2cu^4/4!$ to the Hamiltonian.

\subsubsection{The two-loop diagram} Let us now consider the two-loop diagram
with two vertices (Fig.~2). 
\begin{figure}[h]\centering
\includegraphics{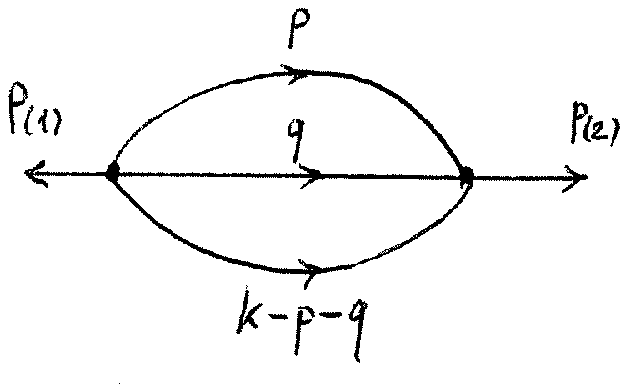}
\caption{}
\end{figure}
To it the following integral corresponds:
\begin{equation}
\int\frac{\widetilde\chi(p_{(1)}+k)\widetilde\chi(p_{(2)}-k)}
{(p^2-m^2)(q^2-m^2)((k-p-q)^2-m^2)}dpdqdk,
\end{equation}
in which we have omitted for shortness the constant before the integral, the symbols $\PV$
and the factor $\prod_{i=1}^2\tilde u(p_{(i)})dp_{(i)}$.
The integral is divergent for large $p,q$. To make it convergent, one can, for instance,
subtract from the fraction
\begin{equation}
\frac1{(p^2-m^2)(q^2-m^2)((k-p-q)^2-m^2)}
\end{equation}
its Taylor polynomial with respect to $k$ at $k=0$ of the second order, i.~e. the terms
of the zeroth, first and second order in the Taylor development. Then the remainder
will be an integral of partial derivatives with respect to $k$ of the third order,
and it is not difficult to see that it would give the convergent integral instead of
(106). But in this process in the numerator the integrals
\begin{equation}
\begin{aligned}{}
&\int k_i\widetilde\chi(p_{(1)}+k)\widetilde\chi(p_{(2)}-k)dk,\\
&\int k_ik_j\widetilde\chi(p_{(1)}+k)\widetilde\chi(p_{(2)}-k)dk,
\end{aligned}
\end{equation}
will occur, which are the Fourier transforms with respect to the variable $p_{(1)}+p_{(2)}$
of the expressions $\chi(x)\frac{\partial\chi}{\partial x_i}(x)$
and $\frac{\partial\chi}{\partial x_i}(x)\frac{\partial\chi}{\partial x_j}(x)$.
And these expressions are not defined as distributions.
Here the problem is that the characteristic function $\chi(x)$
is not differentiable.

Thus, we see that our program of defining dynamical evolution in quantum field theory
fails on the two-loop diagram.

\subsubsection{Dynamical evolution in the quasiclassical approximation}
However, in the one-loop approximation, i.~e., in the sum over the diagrams with no more than
one loop, the program of defining dynamical evolution works well. We come to the following
theorem.
\medskip

{\bf Theorem.} In the $\varphi^4$ model of quantum field theory in four dimensional
space-time the dynamical evolution
exists in the one-loop approximation of perturbation theory.

\medskip

This theorem means that to each pair of space-like surfaces $\calC_1$,
$\calC_2$ one can assign, with the help of the subtraction procedure, an element of the
Weyl algebra $W_0$ of the type
\begin{equation}
P_{\calC_1,\calC_2}(u(\cdot,\cdot))=e^{iS(u(\cdot,\cdot))/h}a(u(\cdot,\cdot)),
\end{equation}
so that conjugation by the element $e^{iS(u(\cdot,\cdot))/h}$ in the Weyl algebra
yields, up to $O(h)$, the evolution operator of the classical field equation (99), and
\begin{equation}
P_{\calC_1,\calC_3}=P_{\calC_2,\calC_3}*P_{\calC_1,\calC_2}+o(h).
\end{equation}
Besides that, as the surface $\calC_1$ tends to $t=-\infty$ and the surface $\calC_2$
to $t=\infty$, the element $P_{\calC_1,\calC_2}$ tends to the $S$-matrix up to $o(h)$
(see the next Subsection).

Proof of this theorem is based on the fact that the only one-loop diagrams
giving divergent integrals are the diagrams containing the diagram
``fish'' from 2.4.7.
Making the same subtraction procedure with them as with the ``fish'' diagram,
we shall obtain the required element $P_{\calC_1,\calC_2}$.

This theorem is in accordance with the results from the book [2] by Maslov and Shvedov,
who constructed complex germ in quantum field theory using the Bogolyubov $S$-matrix
(regarding this $S$-matrix see the next Subsection).

\subsubsection{The scattering matrix}

Thus, for the two-loop diagram the subtraction program meets the difficulty that
the characteristic function $\chi(x)$ of the strip is not differentiable. Let us slightly
change the viewpoint on the dynamical evolution, and let us look not for a family of
isomorphisms of the Weyl algebras of space-like surfaces, related with the integral
(95), but for {\it one} element of the Weyl algebra, playing the role of the evolution
in the whole space-time and related with the integral
\begin{equation}
T\exp\int_{-\infty}^{\infty}\!\int\frac1{ih}g(t,\x)u(t,\x)^4/4!\,dtd\x,
\end{equation}
where $g(t,\x)$ is a smooth function, say, with compact support. In other words,
let us change the function $g\chi(x)$ in our considerations by a differentiable function
$g(x)$, and consider the Lagrangian
\begin{equation}
F(x^\mu,u,u_{x^\mu})=\frac12u_{x^\mu}u_{x_\mu}-\frac{m^2}2u^2-\frac{g(x)}{4!}u^4.
\end{equation}
The perturbation series for this Lagrangian is given by the integral (111).

If we develop this integral according to the rules from 2.4.3 and 2.4.6, i.~e.
develop the summand
\begin{equation}
\int\frac1{(ih)^N4!^N N!}g(x_{(1)})\ldots g(x_{(N)})Tu(x_{(1)})^4*\ldots*u(x_{(N)})^4
\prod dx_{(i)},
\end{equation}
where integration goes over the whole space-time, then the obtained integrals,
related with the Feynman diagrams, will exactly coincide with the Feynman integrals
from Bogolyubov--Shirkov's book [13], with the only difference: instead of function (98),
in the Feynman integrals the propagator
\begin{equation}
\widetilde D_c(p)=\frac1{p^2-m^2+i\varepsilon}.
\end{equation}
stands. This propagator differs from $\widetilde D(p)$ by a multiple of the delta-function
$\delta(p^2-m^2)$,
hence this difference does not affect on the divergences at large momenta. Hence we can
apply to our integrals the subtraction procedure from the Bogolyubov--Shirkov's book
(the Bogolyubov--Parasyuk theorem). In fact, we have already begun to apply it in
2.4.7, 2.4.8.

Having applied this procedure, we will obtain an element $P(g)$ of the Weyl algebra $W_0$,
which is a formal series over the powers of the function
$g(x)$ and which is defined not uniquely, but only up to adding finite terms to the Lagrangian.
Conjugation by the element $P(g)$ in the Weyl algebra gives, up to
$O(h)$, the perturbation series for the evolution operator of the classical field equation
\begin{equation}
\Box u(x)-m^2u(x)=g(x)u^3(x)/3!
\end{equation}
from $t=-\infty$ to $t=\infty$.

Further, consider the operator in the Fock space corresponding to the element $P(g)$
(see 2.3.3). Denote it by $S(g)$. We state that the operator $S(g)$ is exactly the
$S$-matrix constructed in the book [13] by Bogolyubov and Shirkov. This $S$-matrix
is obtained from the element $P(g)$ by the change everywhere of function $\widetilde D(p)$ by
the Feynman propagator $\widetilde D_c(p)$, the $*$-product of functionals
by the composition of operators, and the usual product of functionals (for example, $u^4(x)$)
by the normally ordered product of operators. Indeed, by the Wick theorems (see
[13]), operations with products of functionals in the Weyl algebra, such as
$*$-products with the sign $T$, exactly correspond to operations
with normally ordered products of operators in the Fock space, with the only difference:
the function $D(x)$ should be replaced in these formulas by the function $D_c(x)$.
And the subtraction procedure in the Weyl algebra exactly goes to
the subtraction procedure for operators in the Fock space.

Thus, the operator $S(g)$ is the {\it Bogolyubov $S$-matrix}, or
the {\it scattering matrix}. This $S$-matrix satisfies the
{\it Lorentz invariance, unitarity and causality} conditions. In the Weyl algebra
these conditions go to the corresponding conditions for the element $P(g)$.
The Lorentz invariance condition is obvious:
\begin{equation}
LP(L^{-1}g)=P(g)
\end{equation}
for a Lorentz transformation $L$. The unitarity condition means that
\begin{equation}
P(g)*\overline{P(g)}=1.
\end{equation}
Finally, the causality condition states that for two functions $g_1(x)$ and $g_2(x)$,
coinciding for $t\le t_0$, the element $P(g_1)*P(g_2)^{-1}$ does not depend on the
behavior of the functions $g_1$, $g_2$ for $t<t_0$. These conditions are the
natural substitutes of the conditions for the dynamical evolution. In particular,
the causality condition is the natural substitute of the condition of dependence
of the evolution operator of a linear differential equation on the coefficient
functions of this equation. In Bogolyubov--Shirkov's book [13] it is shown that
these conditions define the
$S$-matrix uniquely, up to adding finite terms to the Lagrangian. Hence it is
natural to postulate the existence of the elements $P(g)$ and $S(g)$
outside of the framework of perturbation theory.

The physical sense of the $S$-matrix is the following. Assume that, as the function
$g(x)$, remaining a function with increasing compact support, tends to the
constant function $g=\const$ (the {\it adiabatic interaction switch off}),
the elements $S(g)$ and $P(g)$ tend to some elements $S$ (the physical $S$-matrix)
and $P$.
Then the square of the absolute value of the matrix element (86) (with $\Phi=P$)
of the physical $S$-matrix is the density of probability of the event that
$N$ colliding particles, flying before collision with 4-momenta $p_{(1)}$, $\ldots$,
$p_{(N)}$, after collision turn into $N'$ particles flying away with 4-momenta
$p'_{(1)}$, $\ldots$, $p'_{(N')}$.

Up to $o(h)$ the element $P$ gives the operator $P_{\calC_1,\calC_2}$
of quasiclassical dynamical evolution from $\calC_1$: $t=-\infty$ to
$\calC_2$: $t=\infty$, see 2.4.9. The definition of dynamical evolution (up to $o(h)$),
suitable outside perturbation theory, can be found in 2.4.4.

Let us also comment on the non-uniqueness of $S$-matrix. Actually it depends
not only on the initial Lagrangian, but also on effective parameters, for example,
on effective mass and effective coupling constant, which are computed from the $S$-matrix
and which already define it uniquely. And any change of effective parameters is equivalent
to certain change of parameters of initial Lagrangian.

In a similar manner one constructs the apparatus of the operator and scalar Green functions,
with the help of the Lagrangian
\begin{equation}
F(x^\mu,u,u_{x^\mu})=\frac12u_{x^\mu}u_{x_\mu}-\frac{m^2}2u^2-\frac{g(x)}{4!}u^4-\jj(x)u.
\end{equation}
Their construction, including the subtraction procedure, does not yield new difficulties,
and is made similarly to what is done in the Bogolyubov--Shirkov's book.

Note that the above constructed apparatus of the $S$-matrix and conditions on it
are analogous to the scattering theory in the theory of partial differential equations,
where, given the coefficient functions of the equation, one is required to determine
the properties of the evolution operator from $t=-\infty$
to $t=\infty$.

\end{document}